\documentclass[11pt,a4paper]{article}
\usepackage{jheppub}
\usepackage[T1]{fontenc}
\usepackage{bm,mathtools,booktabs,mathrsfs}

\newcommand{\Tr}{\operatorname{Tr}}
\newcommand{\tr}{\operatorname{tr}}
\newcommand{\im}{\operatorname{im}}
\newcommand{\rank}{\operatorname{rank}}
\newcommand{\cyc}{\operatorname{cyc}}
\newcommand{\YM}{\mathrm{YM}}
\newcommand{\ee}{\epsilon}
\newcommand{\scrS}{\mathscr{S}}
\newcommand{\Ds}{D_s}

\title{A six-gluon obstruction to dimension-independent local color-kinematics duality at one loop}
\author{Ilmo Sung}
\affiliation{Science and Technology Directorate, U.S. Department of Homeland Security, \\ Washington, DC, USA}
\emailAdd{ilmo.sung@hq.dhs.gov}
\abstract{We establish an obstruction to dimension-independent local color-kinematics duality for the one-loop six-gluon amplitude in pure Yang-Mills theory. The maximal and fivefold cuts admit a polynomial solution, but no allowed contact correction reproduces a required fourfold cut. We consider parity-even Lorentz-polynomial numerators, multilinear in arbitrary transverse external polarizations, with conventional quadratic propagators, global graph identities, and pointwise formal-color cut equality. No dimension-specific Gram identities are imposed. In the sector containing one polarization inner product, the higher cuts determine the complete remaining contact freedom. Every such correction obeys an exact coefficient relation that the required physical residual violates, independently of the chosen higher-cut solution. The violation is proportional to the transverse internal vector-state count and is nonzero at every nonzero value of its algebraic continuation. Momentum homogeneity extends the degree-six contradiction to all finite polynomial degrees. An all-multiplicity polynomial family retaining these representation conditions is therefore excluded.}
\hypersetup{
 pdftitle={A six-gluon obstruction to dimension-independent local color-kinematics duality at one loop},
 pdfauthor={Ilmo Sung},
 pdfsubject={One-loop pure Yang-Mills numerator representations},
 pdfkeywords={color-kinematics duality, Yang-Mills, generalized unitarity, six gluons}
}

\let\jheptableofcontents\tableofcontents
\renewcommand{\tableofcontents}{{\small\jheptableofcontents}}

\makeatletter
\renewcommand{\ps@myplain}{%
  \pagenumbering{arabic}%
  \renewcommand{\@oddfoot}{\hfill\thepage\hfill}%
  \renewcommand{\@evenfoot}{\hfill\thepage\hfill}%
  \renewcommand{\@oddhead}{}%
  \renewcommand{\@evenhead}{}%
}
\let\ps@plain\ps@myplain
\makeatother

\begin{document}
\maketitle
\flushbottom

\section{Introduction}
\label{sec:intro}
Local color-kinematics-dual representations of the one-loop pure-Yang-Mills amplitude have been constructed through five external gluons~\cite{Bern2015,Cao2026}. We establish an obstruction at six points in the dimension-independent representation class defined below. For nonzero transverse internal vector-state count, no finite Lorentz-polynomial numerator assignment satisfies all routed graph identities and reproduces the physical cuts. The maximal and fivefold cuts nevertheless admit a polynomial solution: the incompatibility arises when their complete contact freedom is required to reproduce a fourfold cut.

In a cubic-graph representation, color-kinematics duality requires the numerators to obey the antisymmetry and Jacobi identities of the color factors~\cite{BCJ2008,BCJ2010}, with consistent loop-momentum routing. Generalized unitarity fixes the residues obtained by putting internal propagators on shell~\cite{BDDK1994,BCF2005}. A numerator term proportional to an inverse propagator cancels that propagator in the integrand. Such a contact term does not affect a cut of the canceled propagator, but can contribute when it remains uncut. The organization by maximal cuts~\cite{BernMaximal2007} therefore leaves a completion problem: the remaining terms must reproduce lower cuts while preserving polynomial locality and the graph identities.

Our argument resolves this problem in the sector containing one inner product of external polarization vectors. At momentum degree six, the maximal-cut equation determines the complete form of its contact corrections, and the fivefold equations impose their exchange properties. Routed relabelings then correlate different polarization-pair coefficients on the fourfold cut. An explicit linear combination vanishes for every allowed correction but is nonzero on the required physical residual. Changing the particular solution of the higher cuts leaves this mismatch unchanged. Momentum homogeneity shows that any finite polynomial solution would contain a degree-six solution, so the contradiction excludes arbitrary finite polynomial degree.

A related obstruction is known at two loops and four points: the bowtie cut is incompatible with global color-kinematics duality and polynomial locality~\cite{Edison2024}. Local constructions for that amplitude can retain a subset of the off-shell Jacobi relations~\cite{LiYang2024}. The result here instead concerns the full routed system at one loop and six points. Other positive constructions address different state sectors: four-dimensional all-plus and single-minus amplitudes admit color-dual one-loop numerators at arbitrary multiplicity~\cite{Boels2013}, while higher-point supersymmetric numerators involve different internal state sums~\cite{Edison2023}.

We denote the internal vector-state dimension by $\Ds$; a physical gluon state sum contains $\Ds-2$ transverse states. This parameter is distinct from the kinematic dimension. The exclusion holds at each fixed $\Ds\ne2$ in the characteristic-zero scalar-product algebra defined in Section~\ref{sec:conventions}, which also specifies the representation and cut conditions. Section~\ref{sec:contact} determines the higher-cut solutions, and Section~\ref{sec:obstruction} derives the fourfold-cut obstruction and its finite-polynomial consequence. Section~\ref{sec:discussion} interprets the result and compares it with nearby constructions.

\section{Representation space and cut equations}
\label{sec:conventions}
\subsection{Cubic representation}
For six incoming gluons with momenta $k_i$ and polarization vectors $\ee_i$, the stripped numerator assignment is organized by the formal cubic-graph sum
\begin{equation}
 \mathcal I_6^{(1)}(\ell)
 =\sum_{g\in\mathcal G_6^{(1)}}
 \frac{c_gN_g(\ell)}{S_g\prod_{e\in g}p_e^2}.
 \label{eq:integrand}
\end{equation}
Here $\ell$ is the loop momentum, $\mathcal G_6^{(1)}$ contains the connected one-loop cubic graphs, $c_g$ is a formal adjoint-color factor, $S_g$ is the graph symmetry factor, and $p_e$ is the routed momentum on an internal edge. The gauge coupling and common phases are stripped consistently. All propagators are conventional massless quadratic propagators.

We first analyze numerators $N_g$ that are parity-even Lorentz polynomials, homogeneous of total momentum degree six and linear in each external polarization. Every momentum, including $\ell$, has degree one, while polarizations have degree zero. For a one-loop cubic graph with $E$ internal edges and $V$ vertices, $E=V$ and $3V=2E+6$, so there are six of each. Degree six is the corresponding Yang-Mills numerator degree; resolving a quartic vertex introduces an inverse propagator with the same power counting. There is no additional graph-dependent bound on loop momentum. Numerators contain neither kinematic denominators nor external reference vectors. Section~\ref{sec:homogeneity} treats arbitrary finite polynomial degree.

For each routed color relation, the corresponding numerator relation is imposed off shell. For example, a choice of graph orientations gives
\begin{equation}
 c_i-c_j-c_k=0
 \quad\Longrightarrow\quad
 N_i(\ell_i)-N_j(\ell_j)-N_k(\ell_k)=0.
 \label{eq:jacobi}
\end{equation}
The translated loop arguments are fixed by the graph routes. Vertex antisymmetry is required in the same convention. These are identities of the numerators, not identities restricted to the support of a cut.

For a generalized cut $\mathcal C$, the graph residue must equal the product of Yang-Mills tree amplitudes sewn over physical internal gluon states and color indices. Equality is imposed pointwise as an identity of rational functions on the generic on-shell cut variety, before integration. Formal adjoint color imposes graph antisymmetry and Jacobi identities without additional identities of a fixed finite gauge group.

We use the ordered cut equations obtained by resolving this equality in color. The Del Duca-Dixon-Maltoni decomposition~\cite{DDM2000}, with the two cut legs held fixed at each tree corner, produces chains of adjoint generators. Sewing the corners gives ring color tensors. For the cut classes retained below, these tensors are independent: the explicit trace projection in Appendix~\ref{app:color} extracts each ordered tree product $T_{\mathcal C}$ separately. Thus their ordered cut equations follow from full formal-color cut equality and impose no additional requirement on this subsystem.

All graph topologies remain allowed, including triangles, bubbles, external bubbles, and tadpoles. External-bubble terms may contain $1/k_i^2$~\cite{Bern2015}; the formal sum in Eq.~\eqref{eq:integrand} does not assign their singular quotients a literal on-shell value. The retained ordinary cuts contain at least four distinct loop edges in the main argument and at least three in Appendix~\ref{app:independent}. External bubbles and tadpoles have no residue on this subsystem, so their limiting prescription is not used and their numerators are not set to zero.

\subsection{Scalar products and internal states}
The kinematic constraints are
\begin{equation}
 k_i^2=0,\qquad \sum_{i=1}^{6}k_i=0,\qquad
 \ee_i\cdot k_i=0.
 \label{eq:kinematics}
\end{equation}
The polarizations are arbitrary transverse vectors, and $s_{ij}=2k_i\cdot k_j$. No null-polarization condition is imposed. Polarization self-contractions are absent from the component linear in every $\ee_i$.

We work in the scalar-product algebra generated by
\begin{equation}
 (\ell,k_1,\ldots,k_5,\ee_1,\ldots,\ee_6),\qquad
 k_6=-\sum_{i=1}^{5}k_i,
 \label{eq:vectors}
\end{equation}
subject only to Eq.~\eqref{eq:kinematics}. No dimension-specific Gram determinant relations are imposed. Consequently, sectors with different numbers of cross-polarization inner products are independent. Dimension independence refers to this algebra throughout the paper.

The kinematic dimension $D$ and internal vector-state dimension $\Ds$ have distinct roles~\cite{Giele2008}. Generic complex realizations of Eq.~\eqref{eq:vectors} exist for $D\ge12$, and physical internal-state sums have $\Ds\ge D$. Twelve is sufficient and need not be minimal. The operator traces below define the algebraic continuation in $\Ds$. Evaluating that continuation at four does not impose four-dimensional identities on the external scalar products.

At each fixed $\Ds$, numerator coefficients may be arbitrary elements of a characteristic-zero field $\mathbb F$ and are constant in the kinematics. No polynomial or affine dependence of these unknown coefficients on $\Ds$ is required.

In this algebra, at every fixed $\Ds\ne2$, no assignment of parity-even finite Lorentz-polynomial cubic numerators, multilinear in the six transverse external polarizations, satisfies all routed graph identities and the physical formal-color cut equations with conventional quadratic propagators. The derivation first excludes degree six in Sections~\ref{sec:contact} and \ref{sec:obstruction}; the homogeneous projection in Section~\ref{sec:homogeneity} then establishes the stated finite-polynomial result. The displayed obstruction vanishes at $\Ds=2$, where no existence conclusion follows.

\subsection{Covariant master and loop routing}
Let $N(1,2,3,4,5,6;\ell)$ denote the numerator of the hexagon with the indicated external ordering and incoming loop momentum. Jacobi identities express the numerators entering the retained cuts as signed sums of such ordered hexagon numerators. For example,
\begin{align}
 N([1,2],3,4,5,6;\ell)
 &=N(1,2,3,4,5,6;\ell)\notag\\
 &\quad-N(2,1,3,4,5,6;\ell).
 \label{eq:two-root}
\end{align}
The bracket $[1,2]$ denotes a two-gluon tree attached to the loop, with the indicated commutator orientation. All terms use the same incoming loop momentum at that attachment. Repeating this reduction expands every attached tree into ordered hexagon numerators, called master numerators, with the corresponding loop translations and reversals. No division is introduced, so locality and total degree are preserved. Appendix~\ref{app:coordinates} gives the recursive reduction.

It is sufficient to use a covariant master. If a complete numerator assignment exists, its external relabelings solve the relabeled cut and graph equations. Averaging the whole coupled assignment over the external permutation group therefore produces another solution. Characteristic zero permits division by the group order. The hexagon stabilizer is its dihedral group, so the averaged master has routed cyclic and reflection covariance. The average acts on the entire Jacobi-coupled assignment; crossing symmetry of an initial solution is not an additional assumption.

For the reference ordering $(1,2,3,4,5,6)$, we use the loop momenta, inverse propagators, and centered momentum defined by
\begin{align}
 q_i&=\ell+\sum_{j=1}^{i}k_j,\qquad q_0=q_6=\ell,
 \label{eq:qroute}\\
 \rho_i&=q_i^2,\qquad i=0,\ldots,5,
 \label{eq:rhos}\\
 L&=\ell+K,\qquad
 K=\frac{5k_1+4k_2+3k_3+2k_4+k_5}{6}.
 \label{eq:centered}
\end{align}
The edge immediately after external leg $i$ carries $q_i$, so $\rho_i$ is its inverse propagator, with indices understood cyclically. The centered momentum $L$ makes a cyclic rotation of the hexagon a relabeling at fixed $L$; the conventional incoming momentum $\ell$ shifts with that rotation. It is not an additional independent momentum.

A cut is denoted by the cyclic sequence of its external corner words. Thus $12|3|4|5|6$ places gluons 1 and 2 at the same tree corner. A corner with $b$ external gluons is a $(b+2)$-point tree, including its two cut legs. Table~\ref{tab:cuts} specifies the three cut classes used in the main argument. All their external relabelings are necessary equations of the representation.

After the graph reduction, $C_{\mathcal C}[F]$ denotes the cut residue obtained from a master polynomial $F$. In particular, $C_5$ denotes the displayed fivefold map with physical target $T_5$, and $C_4$ denotes the displayed fourfold map with target $T_{\YM}$.

\begin{table}[htbp]
\caption{Cuts used in the main argument, in the reference route of Eq.~\eqref{eq:qroute}. Each listed $\rho_i$ is set to zero. The last column gives the multiplicity of the nontrivial tree corner, or of each corner for the maximal cut.}
\label{tab:cuts}
\centering
\begin{tabular}{lcl}
\toprule
Corner words & Cut indices $i$ & Tree order\\
\midrule
$1|2|3|4|5|6$ & $0,1,2,3,4,5$ & Three-point\\
$12|3|4|5|6$ & $0,2,3,4,5$ & Four-point\\
$123|4|5|6$ & $0,3,4,5$ & Five-point\\
\bottomrule
\end{tabular}

\end{table}

\section{Polynomial freedom compatible with the higher cuts}
\label{sec:contact}
\subsection{Variations preserving the maximal cut}
We first determine the homogeneous freedom without choosing a particular numerator. Let $\Delta$ be a routed, dihedrally covariant master variation of the degree and polarization multidegree specified in Section~\ref{sec:conventions}, whose maximal and fivefold cuts vanish. On the maximal cut only the hexagon contributes, so $\Delta$ vanishes when all six inverse propagators are zero.

The inverse propagators are independent affine-linear coordinates in the six loop scalar products, with the external invariants retained. The inverse transformation is
\begin{align}
 \ell^2&=\rho_0,
 \label{eq:inverse-one}\\
 2\ell\cdot k_i&=\rho_i-\rho_{i-1}
                   -2\sum_{j<i}k_i\cdot k_j,
 \quad i=1,\ldots,5.
 \label{eq:inverse-two}
\end{align}
The maximal cut is therefore a coordinate plane with ideal
\begin{equation}
 \mathfrak I=(\rho_0,\ldots,\rho_5).
 \label{eq:ideal}
\end{equation}
A polynomial that vanishes at the generic points of this coordinate plane belongs to its ideal; therefore $\Delta\in\mathfrak I$.

The polarization monomials are graded by the number $r$ of factors $\ee_i\cdot\ee_j$. Momentum conservation, transversality, external relabeling, and affine loop rerouting preserve $r$. The grades are independent in the stated scalar-product algebra, so an incompatibility in one grade suffices to exclude a full solution. This is a decomposition of a covariant Lorentz polynomial, not a restriction on the external helicities.

For $r=1$, the polarization inner product uses two external polarizations. The remaining four must each contract with a momentum, using four powers of momentum. Total degree six leaves exactly one additional pure momentum scalar product. Membership in Eq.~\eqref{eq:ideal} then gives
\begin{equation}
 \Delta_{r=1}=\sum_{i=0}^{5}\rho_iQ_i.
 \label{eq:complete-contact}
\end{equation}
Each $Q_i$ contains one polarization inner product and four polarization-momentum contractions, with coefficients in $\mathbb F$ and no additional dependence on pure momentum scalar products. Two inverse-propagator factors would exceed degree six in this grade. Terms of other grades cannot alter its independent coefficients.

An individual monomial of the form in Eq.~\eqref{eq:complete-contact}, before imposing the routed symmetries, is
\begin{equation}
 \begin{split}
 &\rho_0(\ee_1\cdot\ee_2)(\ee_3\cdot k_1)(\ee_4\cdot k_1)\\
 &\qquad\times(\ee_5\cdot\ell)(\ee_6\cdot\ell).
 \end{split}
 \label{eq:contact-example}
\end{equation}
Its inverse propagator supplies two powers of momentum, and its four mixed contractions supply the remaining four. This individual monomial does not satisfy the routed symmetry conditions; those conditions constrain linear combinations of such terms.

For each leg, a basis of mixed contractions consists of the five independent quantities
\begin{equation}
 z_{i,c}=\ee_i\cdot p_{i,c},\qquad c=0,\ldots,4.
 \label{eq:z}
\end{equation}
The momentum options $p_{i,c}$ consist of $L$ and four independent external momenta after momentum conservation and transversality; their ordering is given in Table~\ref{tab:coordinates}. Thus each $Q_i$ belongs to
\begin{equation}
 \mathcal B=\operatorname{span}_{\mathbb F}
 \left\{(\ee_i\cdot\ee_j)
          \prod_{a\notin\{i,j\}}z_{a,c_a}\right\},
 \quad i<j.
 \label{eq:rawspace}
\end{equation}
Its $15\cdot5^4=9375$ monomials exhaust this degree and polarization sector.

\subsection{Homogeneous fivefold constraints}
Let $C$ be the cyclic relabeling $1\mapsto2\mapsto\cdots\mapsto6\mapsto1$ at fixed $L$. It sends $\rho_i$ to $\rho_{i+1}$, with indices modulo six. The reflection fixing gap 1 is
\begin{equation}
 H=(12)(36)(45),\qquad HL=-L,
 \label{eq:H}
\end{equation}
with positive reflection sign for the hexagon. Dihedral covariance organizes Eq.~\eqref{eq:complete-contact} into the cyclic sum
\begin{equation}
 \Delta_{r=1}=\scrS(Q),\qquad
 \scrS(Q)=\sum_{j=0}^{5}C^j(\rho_1Q),\qquad HQ=Q.
 \label{eq:section}
\end{equation}
The coefficients of the independent $\rho_i$ fix this expression uniquely.

On $12|3|4|5|6$, the restriction of the reference master is $\rho_1Q$. The corner has two planar cubic trees, whose commutator expansion assigns weights $1/\rho_1+1/s_{12}$ and $-1/s_{12}$ to the words $123456$ and $213456$. The exchange $\tau=(12)$ holds the incoming $\ell$ fixed, so its action on $L$ is
\begin{equation}
 \tau L=L+\frac{k_2-k_1}{6}.
 \label{eq:tau}
\end{equation}
On the fivefold cut, this exchange gives
\begin{equation}
 \tau\rho_1=-s_{12}-\rho_1.
 \label{eq:swaprho}
\end{equation}
Substituting these two words into the homogeneous cut equation gives
\begin{align}
 C_5[\Delta_{r=1}]
 &=\left(\frac1{\rho_1}+\frac1{s_{12}}\right)\rho_1Q
   -\frac{\tau\rho_1}{s_{12}}\tau Q\notag\\
 &=-\frac{\tau\rho_1}{s_{12}}(Q+\tau Q).
 \label{eq:homogeneous-five}
\end{align}
Its generic vanishing requires $\tau Q=-Q$. Since $Q$ has no pure-momentum scalar-product dependence, this is a polynomial identity in the mixed coordinates, not merely a constraint at a scalar specialization.

Thus the coefficient $Q$ at gap 1 must be even under the edge reflection $H$ and odd under the adjacent-leg exchange $\tau$. On this space the two routed actions are commuting involutions. The projector onto their common required eigenspace is
\begin{equation}
 Q\in\im\Pi,\qquad \Pi=\frac14(1+H)(1-\tau).
 \label{eq:projector}
\end{equation}
Equations~\eqref{eq:section} and \eqref{eq:projector} therefore include every variation in the one-pair sector that can preserve the maximal and fivefold cuts. In particular, they include generalized-gauge and contact transformations within the specified polynomial class.

\subsection{A particular solution of the higher cuts}
To determine the remaining physical cut equation, we choose a polynomial $N_1$ with the required maximal and fivefold cuts. Its construction uses physical-state traces~\cite{Edison2023,Cao2026}.

We denote the identity operator by $I$ and use the outer-product convention $(u\otimes v)z=u(v\cdot z)$. The single-gluon vertex operator is
\begin{equation}
 \begin{split}
 V(q,r;e)={}&e\cdot(q+r)I+(q-2r)\otimes e\\
            &+e\otimes(r-2q),\qquad r=q+k.
 \end{split}
 \label{eq:vertex}
\end{equation}
When $k^2=e\cdot k=0$, contraction gives
\begin{align}
 V(q,r;e)r&=(e\cdot q)q-q^2e,
 \label{eq:vertex-right}\\
 V(q,r;e)^\dagger q&=(e\cdot q)r-r^2e.
 \label{eq:vertex-left}
\end{align}
The adjoint is defined by the Lorentz bilinear form, without complex conjugation of the kinematics.

On the maximal cut, these identities map the null line and its orthogonal hyperplane at one endpoint to the corresponding spaces at the other. Physical states lie in $q_i^\perp/\langle q_i\rangle$, the space of transverse polarizations modulo the gauge direction $q_i$. The trace on this quotient equals the full vector trace minus the contributions of the null line and of the one-dimensional quotient by its orthogonal hyperplane. Both contributions carry the eigenvalue $e\cdot q$ at each vertex. With $V_i=V(q_{i-1},q_i;\ee_i)$, the maximal-cut polynomial is therefore
\begin{equation}
 N_0=\Tr_{\Ds}(V_1V_2V_3V_4V_5V_6)
           -2\prod_{i=1}^{6}(\ee_i\cdot q_{i-1}).
 \label{eq:N0}
\end{equation}
The subtraction removes the two nonphysical contributions. Trace cyclicity and transposition give routed dihedral covariance.

For adjacent legs, the contact operator is
\begin{equation}
 C_{ij}=\ee_i\otimes\ee_j+\ee_j\otimes\ee_i
                   -2(\ee_i\cdot\ee_j)I.
 \label{eq:contactoperator}
\end{equation}
On $12|3|4|5|6$, we use $\rho=\rho_1$, $\bar\rho=(\ell+k_2)^2$, and $s=s_{12}$. The cut conditions then imply $\rho+\bar\rho+s=0$. The four-point corner identity derived in Appendix~\ref{app:operators}, including its physical-state subtraction, gives
\begin{equation}
 C_5[N_0]-T_5
   =\frac{\bar\rho}{s}\Tr_{\Ds}(C_{12}V_3V_4V_5V_6).
 \label{eq:fivefold-mismatch}
\end{equation}
The addition of $\rho\Tr_{\Ds}(C_{12}V_3V_4V_5V_6)/2$ and its exchanged counterpart cancels this mismatch under the two-word cut map. Applying the correction at every adjacent gap yields
\begin{equation}
 N_1=N_0+\frac12\sum_{i=1}^{6}
        \rho_i\Tr_{\Ds}(C_{i,i+1}U_{i,i+1}),
 \label{eq:N1}
\end{equation}
where $U_{i,i+1}$ is the ordered product of the remaining four $V$ operators, starting immediately after the pair. Indices are cyclic, with $\rho_6=\rho_0$.

The polynomial $N_1$ has the stipulated degree, polarization multidegree, and routed covariance, and matches all maximal and generic fivefold cuts. For any hypothetical full solution with master $N$, the difference $\Delta=N-N_1$ therefore has precisely the homogeneous properties derived above. Its one-pair component must be $\scrS(Q)$ with $Q\in\im\Pi$.

\section{The incompatible fourfold cut}
\label{sec:obstruction}
\subsection{Image of the homogeneous contact space}
The remaining necessary equation on $123|4|5|6$ is
\begin{equation}
 C_4[\Delta_{r=1}]
       =\bigl(T_{\YM}-C_4[N_1]\bigr)_{r=1}.
 \label{eq:remaining-cut}
\end{equation}
The left side ranges over the complete homogeneous space of Section~\ref{sec:contact}. The right side is fixed by the Yang-Mills state sum and the particular polynomial $N_1$.

On this cut, $\rho_0=\rho_3=\rho_4=\rho_5=0$. The remaining scalar coordinates are denoted by
\begin{equation}
 \begin{gathered}
 x=\rho_1,\qquad y=\rho_2,\qquad
 a=s_{12},\quad b=s_{23},\quad c=s_{13},\\
 S=a+b+c.
 \end{gathered}
 \label{eq:fourfold-variables}
\end{equation}
The homogeneous master restricts to $xQ+yCQ$. The nontrivial corner is a five-point tree with three external gluons and two cut legs. Its five planar cubic trees give
\begin{equation}
 C_4[F]=\sum_{\sigma\in S_3}m_\sigma
                       F_{\sigma(1)\sigma(2)\sigma(3)456}(\ell).
 \label{eq:fourfold-kernel}
\end{equation}
Every word starts from the same incoming $\ell$. The coefficients $m_\sigma$ contain the uncut loop and tree propagators and the commutator signs. For example,
\begin{align}
 m_{123}&=\frac1{xy}+\frac1{ay}+\frac1{bx}
                       +\frac1{aS}+\frac1{bS},
 \label{eq:m123}\\
 m_{213}&=-\frac{S+y}{ayS}.
 \label{eq:m213}
\end{align}
The compact five-tree commutator expansion in Eq.~\eqref{eq:five-trees} derives these coefficients directly; Table~\ref{tab:weights} lists all six.

The exchange conditions from the fivefold cuts pair these words. For the first free inverse propagator, the exchanged value is $x_2=(\ell+k_2)^2=y-x-a$. The exchange condition $\tau Q=-Q$ gives the paired coefficient
\begin{equation}
 m_{123}x-m_{213}(y-x-a)=f,
 \label{eq:representative-pairing}
\end{equation}
where
\begin{equation}
 f=\frac{a^2+ab+ac+ax+b^2+bc+by}{ab(a+b+c)}.
 \label{eq:f}
\end{equation}
The remaining first-edge pairings and the second-edge pairings, using the oddness of $CQ$ under exchange of legs 2 and 3, give the same factor with the appropriate signs. Their explicit identities in Appendix~\ref{app:certificate} yield
\begin{equation}
 C_4[\scrS(Q)]=f\,\mathcal UQ,\qquad
 \mathcal U=\cyc_{123}(1+C).
 \label{eq:fourfold-map}
\end{equation}
Here $\cyc_{123}$ is the sum of the three cyclic permutations of the corner labels at fixed incoming $\ell$. It differs from the six-leg action $C$, which fixes $L$.

Equation~\eqref{eq:fourfold-map} holds over the rational function field in $a,b,c,x,y$. The free loop poles cancel in the paired coefficients. The remaining denominators $a$, $b$, and $S$ are physical tree propagators; they do not introduce nonpolynomial factors into a graph numerator.

\subsection{A necessary relation and its violation}
The normalized physical residual is
\begin{equation}
 \mathcal R(s,\rho;\Ds)
       =\frac{\bigl(T_{\YM}-C_4[N_1]\bigr)_{r=1}}{f}.
 \label{eq:residual-def}
\end{equation}
Here $s$ and $\rho$ denote the external and loop scalar coordinates. By Eqs.~\eqref{eq:projector}, \eqref{eq:remaining-cut}, and \eqref{eq:fourfold-map}, a completion would require $\mathcal R=\mathcal UQ$ with $Q\in\im\Pi$ wherever $f$ and the cut denominators are nonzero.

Any linear combination of fourfold-cut coefficients that vanishes for every allowed contact contribution must also vanish for the cut mismatch that those contributions are required to reproduce. We use one such weighted combination, denoted by $w$.

The monomials of $\mathcal B$ are ordered by their polarization pair and four mixed-coordinate choices, as specified in Appendix~\ref{app:certificate}. The integer weights defining $w$ have the 96 nonzero components listed in Table~\ref{tab:certificate}. The residual at the regular rational scalar point specified below is denoted by $\mathcal R_*(\Ds)$. The two exact coefficient identities are
\begin{align}
 w\mathcal U\Pi m&=0
       \quad\text{for every }m\in\mathcal B,
 \label{eq:annihilator}\\
 w\mathcal R_*(\Ds)&=-48(\Ds-2).
 \label{eq:physical-obstruction}
\end{align}
The first relation holds for every permitted homogeneous correction; the second evaluates the residual after subtracting $C_4[N_1]$ and dividing by $f$.

In the coordinates of Table~\ref{tab:coordinates}, the support of $w$ involves the polarization pairs $(1,2)$, $(1,4)$, $(1,5)$, $(4,5)$, and $(4,6)$. It links contractions within the three-gluon corner, between that corner and the spectator legs, and among the spectator legs. Restricting $w$ to any one of these five groups gives a nonzero action on $\mathcal U\Pi$; their sum vanishes on the complete space. The relation thus tests correlations among polarization structures, whose cut contributions cannot be adjusted independently under the routed symmetries. This decomposition refers to the stated coordinates; the obstruction is the combined nonzero pairing.

\subsection{Exact evaluation and conclusion}
The homogeneous identity is checked directly on the full monomial space. Expanding $4\mathcal U\Pi=\cyc_{123}(1+C)(1+H)(1-\tau)$ gives 24 signed affine actions. Each action relabels the polarization pair, transforms the four mixed contractions with the prescribed loop shift, and reduces them to the same coordinate dictionary. Multiplication by $6^4$ clears all routing denominators. Pairing with $w$ gives zero on each of the 9375 generators, establishing Eq.~\eqref{eq:annihilator} by integer arithmetic. No numerical rank or selection of a smaller basis enters this identity.

The physical target is computed from ordered Yang-Mills trees with both cubic and quartic interactions. Appendix~\ref{app:operators} specifies the Berends-Giele recursion~\cite{BerendsGiele1988}, endpoint orientations, and physical sewing prescription. The single-gluon operator and the four-point corner fix the normalization relative to $N_1$.

The dependence on $\Ds$ follows from the operator trace. Products of an identity term and outer products have the form
\begin{equation}
 M=\alpha I+\sum_\nu u_\nu\otimes v_\nu.
 \label{eq:operator-decomp}
\end{equation}
An outer product has trace $v_\nu\cdot u_\nu$, so evaluation in the twelve-vector formal basis gives
\begin{equation}
 \Tr_{\Ds}M=\tr_{12}M+(\Ds-12)\alpha.
 \label{eq:state-trace}
\end{equation}
The two nonphysical trace contributions are subtracted as in Eq.~\eqref{eq:N0}. This determines the affine state-count dependence algebraically, without interpolation or a restriction on hypothetical numerator coefficients. No inverse Gram matrix is required.

A rational-function identity valid on the generic cut must hold at every specialization where its denominators are nonzero. It is therefore sufficient to test the necessary relation at
\begin{equation}
 \begin{aligned}
 &(s_{12},s_{13},s_{14},s_{15},s_{23},s_{24},s_{25},s_{34},s_{35})\\
 &\hspace{10pt}=(2,5,7,11,3,13,17,19,23),\\
 &(\rho_0,\ldots,\rho_5)=(0,29,31,0,0,0).
 \end{aligned}
 \label{eq:construction-point}
\end{equation}
Here $f=13/4$, and every uncut tree denominator and temporary reference denominator used in the evaluation is nonzero.

We isolate the coefficient of $(\ee_i\cdot\ee_j)\prod_{a\notin\{i,j\}}z_{a,c_a}$ by setting that polarization-pair coordinate to one, all other cross-polarization coordinates to zero, and all mixed coordinates of the paired legs to zero. For each unpaired leg, the selected mixed coordinate is set to one and the other four to zero. Multilinearity isolates exactly the specified coefficient in the scalar-product algebra. The tree recursion and Eq.~\eqref{eq:state-trace} then determine its value at $\Ds=12$ and its slope.

Table~\ref{tab:certificate} gives all 96 supported coefficients. Their exact weighted sums are
\begin{equation}
 w\mathcal R_*(12)=-480,\qquad
 w\,\partial_{\Ds}\mathcal R_*=-48.
 \label{eq:physical-pairings}
\end{equation}
Thus $w\mathcal R_*(\Ds)=-480-48(\Ds-12)$, establishing Eq.~\eqref{eq:physical-obstruction}. Any full solution would instead require $w\mathcal R_*(\Ds)=0$, and hence
\begin{equation}
 0=-48(\Ds-2).
 \label{eq:contradiction}
\end{equation}
This is impossible in characteristic zero at fixed $\Ds\ne2$. Equation~\eqref{eq:complete-contact} includes every polynomial direction that can change this coefficient sector, and the other polarization grades cannot affect it. The specified six-gluon numerator representation therefore does not exist.

Changing the polynomial representative of the higher-cut solution changes the normalized residual only by an element annihilated by $w$. Indeed, if $N_1'$ is another polynomial master with the same degree, covariance, and maximal and fivefold cuts, its difference from $N_1$ is an allowed homogeneous variation. Hence
\begin{equation}
 w\!\left[
 \frac{\bigl(C_4[N_1']-C_4[N_1]\bigr)_{r=1}}{f}
 \right]=0.
 \label{eq:lift-independence}
\end{equation}
Replacing $N_1$ by $N_1'$ cannot change Eq.~\eqref{eq:physical-obstruction}. At $\Ds=2$, the displayed obstruction vanishes without establishing a full representation.

The state-count slope also has a scalar interpretation. An added internal polarization orthogonal to all momenta and external polarizations sees only the identity component of each corner operator and propagates as one real adjoint scalar in dimensional reduction~\cite{Giele2008}. If $T_{\mathrm{scalar}}$ denotes its cut contribution in the same normalization, the residual satisfies
\begin{equation}
 \partial_{\Ds}\mathcal R
 =\frac{\bigl(T_{\mathrm{scalar}}
       -C_4[\partial_{\Ds}N_1]\bigr)_{r=1}}{f}.
 \label{eq:scalar-residual}
\end{equation}
Appendix~\ref{app:scalar} evaluates both terms by scalar-line recursion and reproduces the supported slopes. The factor $\Ds-2$ also uses the constant term in Eq.~\eqref{eq:physical-pairings}.

Appendix~\ref{app:independent} describes an additional exact reconstruction using a different maximal-cut polynomial and a specialization of the external polarizations. Every allowed specialized correction lies in an enlarged polynomial space. Exact rational elimination of its cut equations yields the same incompatibility. External Ward identities, reference-independent transverse-state sewing, routed symmetries, and a published four-point numerator provide further checks. Appendix~\ref{app:certificate} also reports the evaluation of the primary residual at a second regular rational point.

\subsection{Momentum homogeneity and finite polynomials}
\label{sec:homogeneity}
The degree-six obstruction also excludes finite polynomial numerators of arbitrary degree, provided that the remaining representation conditions are unchanged. Suppose that such an assignment exists. Each numerator admits a finite decomposition into homogeneous momentum components,
\begin{equation}
 N_g=\sum_d N_g^{(d)}.
 \label{eq:degree-decomposition}
\end{equation}
Under a uniform rescaling of all external and loop momenta by a nonzero parameter $t$, with the polarizations and coefficient parameters held fixed, the kinematic relations remain homogeneous and the routed momentum translations remain linear. The graph antisymmetry and Jacobi identities therefore hold separately at each momentum degree.

A connected one-loop six-point cubic graph has six internal edges. On a cut with $m$ distinct loop edges, $6-m$ propagators remain, so
\begin{equation}
 C_m[N^{(d)}](t\ell,tk)
 =t^{d-2(6-m)}C_m[N^{(d)}](\ell,k).
 \label{eq:graph-homogeneity}
\end{equation}
Here $C_m$ denotes any retained cut with $m$ edges. A corner with $b_j$ external gluons is a $(b_j+2)$-point stripped Yang-Mills tree of momentum degree $2-b_j$. Physical polarization sewing has degree zero. Since $\sum_{j=1}^m b_j=6$, the cut target obeys
\begin{equation}
 T_m(t\ell,tk)=t^{2m-6}T_m(\ell,k).
 \label{eq:target-homogeneity}
\end{equation}
The cut equation for the hypothetical assignment consequently requires
\begin{equation}
 \sum_d t^{d-6}C_m[N^{(d)}]=T_m
 \label{eq:scaling-cut-equation}
\end{equation}
for arbitrary nonzero $t$. This equality is an identity of finite Laurent polynomials in $t$ over a field of characteristic zero. Comparing coefficients therefore gives
\begin{equation}
 C_m[N^{(6)}]=T_m,\qquad
 C_m[N^{(d)}]=0\quad(d\ne6).
 \label{eq:degree-six-projection}
\end{equation}
The degree-six components would thus satisfy the same graph identities and every necessary cut used in the contradiction. They are excluded by Eq.~\eqref{eq:contradiction}.

The argument imposes no upper bound on the finite degree. Constant dimensionful coefficients remain fixed under scaling and do not alter the projection.

\section{Discussion}
\label{sec:discussion}
The obstruction concerns the compatibility of contact terms with global graph identities. The maximal and fivefold cuts admit a polynomial master, but the complete remaining freedom cannot supply the fourfold residual. The exact functional tests a correlation among polarization structures enforced by routed relabeling. Its nonzero pairing is unchanged by replacing the higher-cut representative, and generalized-gauge transformations within the polynomial class are included in the homogeneous freedom.

The five-point and six-point problems share the ordered five-point tree corner, appearing on $123|4|5$ and $123|4|5|6$, respectively. Their global contact spaces nevertheless differ. A one-pair seed contains three mixed polarization-momentum contractions for the pentagon and four for the hexagon. Reversal changes the orientation of all $n$ cubic vertices and gives sign $(-1)^n$: the corresponding edge-reflection seed is odd at five points and even at six. For the corresponding centered momentum $L_n=n^{-1}\sum_{j=0}^{n-1}q_j$, the adjacent-leg exchange remains odd and shifts $L_n$ by $(k_2-k_1)/n$. These tensor degrees and routed actions distinguish the two completion problems. The explicit five-point solution of Cao et al.~\cite{Cao2026} provides a compatible completion there; at six points the displayed functional excludes the required residual. This comparison does not identify reflection parity alone as the cause of the obstruction.

The scope of the result depends on the simultaneous representation conditions. Partial-fraction forward-limit representations use propagators linear in loop momentum~\cite{HeSchlotterer2017,HeSchlottererZhang2018}, while tensorial Parke-Taylor expansions produce quadratic-propagator representations~\cite{Feng2022,Dong2024}. The explicit Yang-Mills examples in Ref.~\cite{Dong2024} have all-plus or single-minus helicities. The broader construction of Ref.~\cite{Cao2026} requires shifted cyclic compatibility. Formulas organized in effective currents and local multipoint vertices~\cite{XieDu2025,XieDu2026} do not by themselves provide a cubic expansion satisfying every loop-graph Jacobi identity.

Locality in all momentum contractions is also stronger than polynomial dependence on loop momentum. The Yang-Mills-scalar coefficients of Ref.~\cite{Du2026} contain factors $(\ee_p\cdot\ee_q)/(k_p\cdot k_q)$; the construction of Ref.~\cite{Mogull2015} increases the loop-momentum degree while allowing external kinematic poles. For color-dual actions obtained through field redefinitions, any Jacobian contributions required for quantum equivalence must be included in the physical-cut comparison~\cite{Borsten2023}.

Momentum homogeneity extends the degree-six contradiction to all finite polynomial degrees. An all-multiplicity family retaining the stated conditions is therefore excluded because it would contain a six-point member. Dimension-specific Gram identities, nonpolynomial numerators, restricted helicities, fixed-group color identities, altered propagators, or equality only after integration change the existence problem. The calculation determines neither a minimum pole structure nor a sufficient relaxation. Any all-multiplicity one-loop construction must depart from at least one of these representation conditions.

\acknowledgments
The author thanks Ryumi S. for inspiration, Yu-tin Huang for reading the manuscript and providing helpful comments and suggestions, and Damian G. and Chris M. for useful discussions. The views expressed are those of the author and do not necessarily represent the U.S. Department of Homeland Security or the United States Government.

\appendix

\section{Routed identities and scalar-product coordinates}
\label{app:coordinates}
\subsection{Independent coordinates}
A convenient choice of external scalar coordinates is
\begin{equation}
 (s_{12},s_{13},s_{14},s_{15},s_{23},s_{24},s_{25},s_{34},s_{35}).
 \label{eq:external-coordinates}
\end{equation}
Masslessness of $k_6=-\sum_{i=1}^{5}k_i$ gives
\begin{equation}
 s_{45}=-(s_{12}+s_{13}+s_{14}+s_{15}+s_{23}
              +s_{24}+s_{25}+s_{34}+s_{35}).
 \label{eq:s45}
\end{equation}
Products with $k_6$ follow from momentum conservation. Equations~\eqref{eq:inverse-one} and \eqref{eq:inverse-two} supply the inverse change from the six $\rho_i$ to the loop scalar products.

Table~\ref{tab:coordinates} gives the five mixed-coordinate options for every leg. The omitted contraction for each leg $i\le5$ vanishes by transversality. For leg 6, momentum conservation and transversality give
\begin{equation}
 \ee_6\cdot k_5=-\sum_{j=1}^{4}\ee_6\cdot k_j.
 \label{eq:eps6}
\end{equation}
Together with the fifteen distinct $\ee_i\cdot\ee_j$, the nine external invariants, six loop invariants, and thirty mixed contractions give sixty independent coordinates relevant to the multilinear component. Polarization self-contractions never enter this component.
\begin{table}[htbp]
\caption{Momentum options defining $z_{i,c}=\ee_i\cdot p_{i,c}$. Choice indices are zero based. The ordered columns fix the monomial encoding used in Table~\ref{tab:certificate}.}
\label{tab:coordinates}
\centering
\begin{tabular}{cccccc}
\toprule
$i$ & $p_{i,0}$ & $p_{i,1}$ & $p_{i,2}$ & $p_{i,3}$ & $p_{i,4}$\\
\midrule
1 & $L$ & $k_2$ & $k_3$ & $k_4$ & $k_5$\\
2 & $L$ & $k_1$ & $k_3$ & $k_4$ & $k_5$\\
3 & $L$ & $k_1$ & $k_2$ & $k_4$ & $k_5$\\
4 & $L$ & $k_1$ & $k_2$ & $k_3$ & $k_5$\\
5 & $L$ & $k_1$ & $k_2$ & $k_3$ & $k_4$\\
6 & $L$ & $k_1$ & $k_2$ & $k_3$ & $k_4$\\
\bottomrule
\end{tabular}

\end{table}

\subsection{Affine routing actions}
For a permutation $\sigma$ performed at fixed incoming $\ell$, the permuted shift is
\begin{equation}
 K_\sigma=\frac{5k_{\sigma(1)}+4k_{\sigma(2)}+3k_{\sigma(3)}
                      +2k_{\sigma(4)}+k_{\sigma(5)}}{6}.
 \label{eq:Ksigma}
\end{equation}
The centered variable transforms as
\begin{equation}
 L\longmapsto L+K_\sigma-K.
 \label{eq:general-shift}
\end{equation}
All momenta and polarizations are relabeled simultaneously. Momentum conservation and transversality then reduce the transformed contractions to Table~\ref{tab:coordinates}.

The six-leg action $C$ fixes $L$ and sends the incoming loop momentum to $\ell+k_1$. Thus $C\rho_i=\rho_{i+1}$, with cyclic indices. The reflection in Eq.~\eqref{eq:H} acts on the incoming momentum as
\begin{equation}
 H\ell=-\ell-k_1-k_2=-q_2,
 \label{eq:Hroute}
\end{equation}
and $H\rho_1=\rho_1$. The fixed-$\ell$ transposition instead gives Eq.~\eqref{eq:tau}. These actions specify the polynomial substitutions, including the signs of every mixed contraction.

An affine action is specified by its leg permutation, the sign multiplying $L$, and the rational coefficients of its external-momentum shift. Composition means substitution of the outer action into the inner one. Each mixed contraction becomes a linear combination of the five options of its relabeled leg. For the actions used here, these coefficients have denominators dividing six, so a product of four mixed contractions has denominator dividing $6^4$.

\subsection{Rooted-tree reduction and the coupled average}
With the incoming loop momentum at a rooted attachment held fixed, the graph identity gives
\begin{equation}
 N(\ldots,[A,B],\ldots)
 =N(\ldots,A,B,\ldots)-N(\ldots,B,A,\ldots).
 \label{eq:root-jacobi}
\end{equation}
The two-leg example is Eq.~\eqref{eq:two-root}. A further application gives
\begin{align}
 N([[1,2],3],4,5,6;\ell)
 &=N(1,2,3,4,5,6;\ell)\notag\\
 &\quad-N(2,1,3,4,5,6;\ell)\notag\\
 &\quad-N(3,1,2,4,5,6;\ell)\notag\\
 &\quad+N(3,2,1,4,5,6;\ell).
 \label{eq:three-root}
\end{align}
Each word on the right starts from the same incoming momentum at the expanded corner. The cumulative momentum changes within that word determine its propagators and vertex arguments. Repetition of Eq.~\eqref{eq:root-jacobi} terminates when all external leaves are attached individually to the loop. A graph needed on one of the retained cuts therefore reduces to routed master words without division.

The permutation group acts on the entire vector of graph numerators, including their labels and routes. If the graph and cut equations are denoted collectively by $\mathcal L(N)=T$, equivariance implies $\mathcal L(\sigma\cdot N)=T$ for the corresponding covariant target assignment. Hence
\begin{equation}
 \overline N=\frac1{6!}\sum_{\sigma\in S_6}\sigma\cdot N
 \label{eq:average}
\end{equation}
is also a solution. The hexagon component of $\overline N$ is covariant under its dihedral stabilizer. This averaging takes place before selecting a master and preserves every relation coupling different graphs. It establishes the necessity of the master space used in the main text without an assumption about the symmetry of an arbitrary initial solution.

\subsection{Formal-color projection of the cut}
\label{app:color}
Let $a_i$ be the external color index of leg $i$, and let $\mathsf F^a$ denote an adjoint generator. Common factors of the coupling and phases are stripped with the graph orientations of Section~\ref{sec:conventions}. With the two cut legs as fixed endpoints, the Del Duca-Dixon-Maltoni tree decomposition~\cite{DDM2000} gives
\begin{equation}
 \begin{aligned}
 &\mathcal A_5^{b_0b_3}(q_0,1,2,3,-q_3)\\
 &\quad=\sum_{\sigma\in S_3}
 (\mathsf F^{a_{\sigma(1)}}\mathsf F^{a_{\sigma(2)}}
  \mathsf F^{a_{\sigma(3)}})_{b_0b_3}\\
 &\qquad\times A_5(q_0,\sigma(1),\sigma(2),\sigma(3),-q_3).
 \end{aligned}
 \label{eq:ddm-corner}
\end{equation}
The Lorentz states on the two cut legs are understood. Sewing the three single-gluon corners yields
\begin{align}
 T_{\mathcal C}^{\mathrm{color}}&=\sum_{\sigma\in S_3}c_\sigma T_\sigma,
 \label{eq:color-cut}\\
 c_\sigma&=\Tr_{\mathrm{ad}}
 (\mathsf F^{a_{\sigma(1)}}\mathsf F^{a_{\sigma(2)}}
  \mathsf F^{a_{\sigma(3)}}\mathsf F^{a_4}\mathsf F^{a_5}\mathsf F^{a_6}),
 \label{eq:color-ring}
\end{align}
where $\mathcal C=123|4|5|6$ and
\begin{equation}
 \begin{split}
 T_\sigma=\sum_{\text{physical states}}&
 A_5(q_0,\sigma(1),\sigma(2),\sigma(3),-q_3)\\
 &\times A_3(q_3,4,-q_4)A_3(q_4,5,-q_5)\\
 &\times A_3(q_5,6,-q_0).
 \end{split}
 \label{eq:ordered-color-target}
\end{equation}
Thus the reference coefficient $T_{123}$ is the physical target $T_{\YM}$.

The independence needed in Eq.~\eqref{eq:color-cut} can be verified by a trace projection. The adjoint action on generic $N_c\times N_c$ matrices is realized as $\operatorname{ad}_{t_i}=L_{t_i}-R_{t_i}$, where $t_i=t^{a_i}$ is traceless and $L_{t_i}X=t_iX$, $R_{t_i}X=Xt_i$. Expanding a product of six commutator actions gives
\begin{equation}
 \begin{aligned}
 &\Tr_{\mathrm{ad}}(\operatorname{ad}_{t_1}\cdots
                   \operatorname{ad}_{t_6})\\
 &\quad=N_c\bigl[\tr(t_1t_2t_3t_4t_5t_6)
                +\tr(t_6t_5t_4t_3t_2t_1)\bigr]\\
 &\qquad+\text{products of two nonempty traces}.
 \end{aligned}
 \label{eq:adjoint-fundamental-traces}
\end{equation}
The identity matrix is annihilated by every commutator, so taking the trace on the adjoint rather than the full matrix space does not change this expression. The plus sign between the two single traces is the even six-point reversal sign.

In the formal trace algebra, let $\mathcal P_\sigma$ extract the coefficient of $N_c\tr(t_{\sigma(1)}t_{\sigma(2)}t_{\sigma(3)}t_4t_5t_6)$, modulo cyclicity. The words $\sigma(123)456$ occupy distinct cyclic-and-reversal classes, so the projection satisfies
\begin{equation}
 \mathcal P_\sigma(c_\tau)=\delta_{\sigma\tau}.
 \label{eq:color-projection}
\end{equation}
This is coefficient extraction in a formal identity, not a leading-color approximation. It requires no claim about the absence of trace identities at a fixed small value of $N_c$. Applying $\mathcal P_{123}$ to the color-dressed cut equation gives precisely the ordered equation used in Section~\ref{sec:obstruction}.

The fivefold cut has the two ring words $123456$ and $213456$, which are distinguished by the same projection; the maximal cut has a single ring word. The construction also distinguishes the 24 words $\sigma(1234)56$ on the threefold cuts of Appendix~\ref{app:independent}. Full formal-color cut equality therefore implies every ordered equation used by either necessary subsystem. Expanding the graph-side corner in the same color basis selects the five terms in Eq.~\eqref{eq:five-trees}.

\section{Physical-state traces and fivefold-cut identities}
\label{app:operators}
\subsection{The adjacent-pair identity}
For the unresolved pair $(12)$, we use the endpoint momenta
\begin{equation}
 q=q_0,\qquad r=q+k_1+k_2,\qquad q^2=r^2=0,
 \label{eq:pair-momenta}
\end{equation}
and the scalar invariants
\begin{equation}
 \rho=(q+k_1)^2,\qquad \bar\rho=(q+k_2)^2,\qquad s=s_{12}.
 \label{eq:fivefold-rhos}
\end{equation}
The cut implies $\rho+\bar\rho+s=0$. The polarization contraction, two-gluon current, and quartic operator are denoted by
\begin{align}
 h&=\ee_1\cdot\ee_2,\notag\\
 v&=h(k_1-k_2)+2(\ee_1\cdot k_2)\ee_2
                         -2(\ee_2\cdot k_1)\ee_1,
 \label{eq:pair-v}\\
 R&=2\ee_2\otimes\ee_1-\ee_1\otimes\ee_2-hI.
 \label{eq:pair-R}
\end{align}
The ordered product of the two single-gluon operators is
\begin{equation}
 A=V(q,q+k_1;\ee_1)V(q+k_1,r;\ee_2),
 \label{eq:pair-A}
\end{equation}
and $A'$ denotes the product with legs 1 and 2 exchanged at fixed $q$. The four-point corner between physical endpoint states is represented by
\begin{equation}
 M=\frac{A}{\rho}-\frac{V(q,r;v)}s+R.
 \label{eq:Mpair}
\end{equation}
A longitudinal endpoint term is immaterial after physical sewing.

For the following identities, we use the scalar abbreviations
\begin{equation}
 \alpha=\ee_1\cdot q,\quad \beta=\ee_2\cdot q,\quad
 \gamma=\ee_1\cdot k_2,\quad \delta=\ee_2\cdot k_1.
 \label{eq:pair-abcd}
\end{equation}
The auxiliary quantities are
\begin{align}
 F&=h(k_1-k_2)-(\beta+2\delta)\ee_1
                         +(\alpha+2\gamma)\ee_2,
 \label{eq:pair-F}\\
 G&=h(k_1-k_2)+(\beta-\delta)\ee_1
                         +(\gamma-\alpha)\ee_2,
 \label{eq:pair-G}\\
 \xi&=\alpha(\beta+\delta)-\beta(\alpha+\gamma)+v\cdot q.
 \label{eq:pair-xi}
\end{align}
Direct multiplication gives
\begin{equation}
 A-A'+V(q,r;v)-sR
       =\bar\rho C_{12}+q\otimes F+G\otimes r,
 \label{eq:first-contact}
\end{equation}
with
\begin{equation}
 F\cdot r=q\cdot G=\xi.
 \label{eq:pair-contract}
\end{equation}
The endpoint identities are
\begin{align}
 Mr&=\lambda q,\qquad M^\dagger q=\lambda r,
 \label{eq:pair-endpoint}\\
 \lambda&=\frac{\alpha(\beta+\delta)}\rho-\frac{v\cdot q}s.
 \label{eq:pair-lambda}
\end{align}
They follow in the formal span of $q,k_1,k_2,\ee_1,\ee_2$ from $k_i^2=0$, $\ee_i\cdot k_i=0$, and $\rho+\bar\rho+s=0$.

The remaining operator chain and its scalar endpoint factor are
\begin{equation}
 U=V_3V_4V_5V_6,\qquad
 g=\prod_{i=3}^{6}(\ee_i\cdot q_{i-1}).
 \label{eq:remaining-chain}
\end{equation}
Equations~\eqref{eq:vertex-right} and \eqref{eq:vertex-left} imply $Uq=gr$ and $U^\dagger r=gq$. The physical trace is therefore
\begin{equation}
 T_5=\Tr_{\Ds}(MU)-2\lambda g.
 \label{eq:T5}
\end{equation}
The two rank-one terms on the right of Eq.~\eqref{eq:first-contact} contribute $2\xi g$ to the metric trace. The same term occurs in the nonphysical trace subtraction and therefore cancels. Using the weights $1/\rho+1/s$ and $-1/s$ then gives Eq.~\eqref{eq:fivefold-mismatch}. The symmetric correction $\rho\Tr_{\Ds}(C_{12}U)/2$ and its exchanged counterpart contribute
\begin{equation}
 \frac12\left[\left(\frac1\rho+\frac1s\right)\rho
                         -\frac{\bar\rho}{s}\right]
       \Tr_{\Ds}(C_{12}U)
   =-\frac{\bar\rho}{s}\Tr_{\Ds}(C_{12}U),
 \label{eq:fivefold-cancel}
\end{equation}
which cancels that mismatch. The cyclic sum in Eq.~\eqref{eq:N1} gives the corresponding correction on every adjacent-pair fivefold cut and vanishes on the maximal cut.

We verified all 25 entries of Eq.~\eqref{eq:first-contact}, its two scalar contractions, and its ten endpoint components by exact symbolic reduction using the stated kinematic constraints, without numerical substitutions for the invariants.

\subsection{Ordered tree recursion}
The physical trees are generated by Berends-Giele recursion~\cite{BerendsGiele1988} with the stripped cubic and quartic currents. For currents $J,K$ carrying incoming momenta $p,q$, the cubic current is
\begin{equation}
 \begin{split}
 \mathcal V_3(p,q;J,K)={}&(J\cdot K)(p-q)\\
 &+\bigl[(p+2q)\cdot J\bigr]K\\
 &-\bigl[(2p+q)\cdot K\bigr]J.
 \end{split}
 \label{eq:BG3}
\end{equation}
For conserved currents, $p\cdot J=q\cdot K=0$, this becomes
\begin{equation}
 \mathcal V_3=(J\cdot K)(p-q)+2(J\cdot q)K-2(K\cdot p)J.
 \label{eq:BG3-transverse}
\end{equation}
The quartic current is
\begin{equation}
 \mathcal V_4(J,K,H)
       =2(J\cdot H)K-(J\cdot K)H-(K\cdot H)J.
 \label{eq:BG4}
\end{equation}
These conventions are consistent with the ordered gluon recursion of Ref.~\cite{CruzMartinez2025}.

For an ordered word $P$, let $p_P$ denote its total momentum. The one-gluon current is $J(i)=\ee_i$, and the current for a word of length at least two satisfies
\begin{equation}
 \begin{split}
 J(P)=\frac1{p_P^2}\Bigg[&\sum_{P=XY}
      \mathcal V_3(p_X,p_Y;J(X),J(Y))\\
 &+\sum_{P=XYZ}\mathcal V_4(J(X),J(Y),J(Z))\Bigg].
 \end{split}
 \label{eq:BGrecursion}
\end{equation}
Every ordered split into nonempty consecutive words is included. The last current is amputated to obtain an on-shell tree amplitude. Thus the recursion includes all cubic and quartic trees, together with their relative signs and propagators.

An open-index recursion gives the corner matrix directly. For a corner with incoming cut momentum $p$ and external word $(1,\ldots,b)$, we use $k_{u:v}=\sum_{i=u}^{v}k_i$ and $J_{u:v}=J(u,\ldots,v)$. The recursion starts from $B_0=I$ and uses the open cubic and quartic operators
\begin{align}
 \mathcal A(p,q;J)
 &=J\otimes(p-q)+(p+2q)\otimes J\notag\\
 &\quad-\bigl[(2p+q)\cdot J\bigr]I,
 \label{eq:open-cubic}\\
 \mathcal W(J,K)&=2K\otimes J-J\otimes K-(J\cdot K)I.
 \label{eq:open-quartic}
\end{align}
For $1\le m\le b$, the recurrence gives the numerator
\begin{equation}
 \begin{split}
 \widetilde B_m={}&\sum_{j=0}^{m-1}B_j
       \mathcal A(p+k_{1:j},k_{j+1:m};J_{j+1:m})\\
 &+\sum_{0\le j<r<m}B_j
       \mathcal W(J_{j+1:r},J_{r+1:m}).
 \end{split}
 \label{eq:open-recursion}
\end{equation}
An empty momentum sum is zero. The intermediate current is $B_m=\widetilde B_m/(p+k_{1:m})^2$ when $m<b$. Amputation gives $B_b=\widetilde B_b$ at the final endpoint, so $B_b$ is the corner matrix.

For one external gluon, this matrix is $-V$ of Eq.~\eqref{eq:vertex}; for two external gluons, it is $M$ of Eq.~\eqref{eq:Mpair}. The six single-gluon factors therefore reproduce Eq.~\eqref{eq:N0}, and the pair corner sewn to four single-gluon factors reproduces Eq.~\eqref{eq:T5}. These equalities fix the relative cut normalizations. In particular, the open-index recursion requires no additional sign assigned solely from the number of cut edges.

\subsection{Physical-state sums}
For a sequence of amputated corner matrices $T_j$ between null momenta $q_j$ and $q_{j+1}$, the endpoint relations take the form
\begin{equation}
 T_jq_{j+1}=\lambda_jq_j,\qquad
 q_j^\flat T_j=\mu_jq_{j+1}^\flat.
 \label{eq:general-endpoints}
\end{equation}
Here $q^\flat$ denotes contraction with the Lorentz metric. If $T_j$ has identity coefficient $\alpha_j$, the physical quotient trace of the closed chain is
\begin{equation}
 \begin{split}
 T_{\rm phys}(\Ds)={}&\tr_{12}(T_1\cdots T_m)
       -\prod_{j=1}^{m}\lambda_j-\prod_{j=1}^{m}\mu_j\\
 &+(\Ds-12)\prod_{j=1}^{m}\alpha_j.
 \end{split}
 \label{eq:general-physical-trace}
\end{equation}
The identity coefficient is propagated algebraically during recursion and multiplication.

For a null cut momentum $q$ and a reference vector $n$ with $q\cdot n\ne0$, the mixed-index physical-polarization completeness tensor is
\begin{equation}
 P(q,n)=-I+\frac{q\otimes n+n\otimes q}{q\cdot n}
                  -\frac{n^2q\otimes q}{(q\cdot n)^2}.
 \label{eq:physical-projector}
\end{equation}
The reference need not be null. Its dependence cancels by the tree Ward identities, and no reference vector enters the numerator space. The negative of Eq.~\eqref{eq:physical-projector} is the idempotent transverse projector, with the sign fixed by the metric and tensor orientations. On the fourfold cut, four such signs multiply to one.

The same cut can be obtained by inserting the idempotent projectors $-P(q_j,n_j)$ with these orientations. We checked their idempotence, annihilation of $q_j$, and trace $D-2$, and verified that the contracted cut agrees with Eq.~\eqref{eq:general-physical-trace}. This tests the physical state sum separately from the quotient-trace evaluation.

\subsection{Scalar contribution to the state-count derivative}
\label{app:scalar}
The identity coefficients of the open cubic and quartic operators in Eqs.~\eqref{eq:open-cubic} and \eqref{eq:open-quartic} are
\begin{equation}
 -\bigl(2p+q\bigr)\cdot J,\qquad -J\cdot K,
 \label{eq:scalar-vertices}
\end{equation}
respectively. They give a scalar-line recursion using the ordinary external-gluon currents $J_{u:v}$. For a corner of length $b$ with incoming momentum $p$, the initial condition is $\beta_0=1$ and the recurrence is
\begin{equation}
 \begin{split}
 \widetilde\beta_m={}&-\sum_{j=0}^{m-1}\beta_j
 (2p+2k_{1:j}+k_{j+1:m})\cdot J_{j+1:m}\\
 &-\sum_{0\le j<r<m}\beta_j
       J_{j+1:r}\cdot J_{r+1:m}.
 \end{split}
 \label{eq:scalar-recursion}
\end{equation}
The intermediate scalar current is $\beta_m=\widetilde\beta_m/(p+k_{1:m})^2$ for $m<b$, while the amputated corner is $\beta_b=\widetilde\beta_b$. This is the identity component of the full corner recursion, obtained without forming its tensor matrix. The product of the $\beta_b$ over the cut corners is $T_{\mathrm{scalar}}$, the contribution of one additional internal polarization orthogonal to all external data. The two nonphysical quotient-trace terms in Eq.~\eqref{eq:general-physical-trace} do not depend on $\Ds$.

For the particular higher-cut solution, the identity coefficient of the single-gluon operator is $\alpha_i=\ee_i\cdot(q_{i-1}+q_i)$. Taking identity coefficients in Eq.~\eqref{eq:N1} gives
\begin{equation}
 \partial_{\Ds}N_1
 =\prod_{i=1}^{6}\alpha_i
 -\sum_{i=1}^{6}\rho_i(\ee_i\cdot\ee_{i+1})
                  \prod_{j\notin\{i,i+1\}}\alpha_j,
 \label{eq:scalar-lift}
\end{equation}
with cyclic indices. Equations~\eqref{eq:scalar-recursion} and \eqref{eq:scalar-lift} evaluate the two terms of Eq.~\eqref{eq:scalar-residual} separately. At both scalar points in Appendix~\ref{app:certificate}, this scalar calculation reproduces all 96 supported state-count slopes and their weighted sum $-48$. It determines the derivative of the residual, not an equality between the full gluon amplitude and a scalar amplitude multiplied by $\Ds-2$.

\subsection{Exact coefficient extraction}
For the primary coefficient table, Gram entries are set directly from Eqs.~\eqref{eq:external-coordinates}, \eqref{eq:inverse-one}, and \eqref{eq:inverse-two}, together with the mixed-coordinate assignments below Eq.~\eqref{eq:construction-point}. The relation between centered and incoming-loop contractions is
\begin{equation}
 \ee_i\cdot\ell=z_{i,0}
       -\frac16\sum_{j=1}^{5}(6-j)\ee_i\cdot k_j.
 \label{eq:ell-eps}
\end{equation}
The operators act on a formal coordinate basis using the Gram matrix only in scalar products and mixed outer products. Singular specializations therefore do not require matrix inversion. Multilinearity excludes all unwanted polarization sectors in the selected coefficient evaluation.

\section{Fourfold coefficients and the exact functional}
\label{app:certificate}
\subsection{The six-word corner kernel}
The reference color projection of Appendix~\ref{app:color} selects five planar cubic trees at the corner on $123|4|5|6$. They have ordered attachments $(1,2,3)$, $([1,2],3)$, $(1,[2,3])$, $[[1,2],3]$, and $[1,[2,3]]$. A rooted subtree contributes its external propagator, while successive loop attachments contribute the intervening uncut loop propagators. We abbreviate $F_{abc}=F(a,b,c,4,5,6;\ell)$, with the same incoming $\ell$ in every term. The sum of the five tree contributions is
\begin{equation}
 \begin{split}
 C_4[F]={}&\frac{F_{123}}{xy}
 +\frac{F_{123}-F_{213}}{ay}
 +\frac{F_{123}-F_{132}}{bx}\\
 &+\frac{F_{123}-F_{213}-F_{312}+F_{321}}{aS}\\
 &+\frac{F_{123}-F_{132}-F_{231}+F_{321}}{bS}.
 \end{split}
 \label{eq:five-trees}
\end{equation}
The five numerators are the corresponding commutator expansions. Collecting the six ordered words gives Table~\ref{tab:weights}. Equivalently, these coefficients form a fixed-ordering row of the five-point biadjoint-scalar double-partial amplitude matrix~\cite{CHY2014}; Eq.~\eqref{eq:five-trees} fixes the signs and loop routing used here.
\begin{table}[htbp]
\caption{Complete scalar kernel of the three-gluon corner, with $S=a+b+c$. Each master word starts from the same incoming $\ell$. The entries include all five planar cubic trees and their commutator signs.}
\label{tab:weights}
\centering
\begin{tabular}{cl}
\toprule
Word $\sigma$ & $m_\sigma$\\
\midrule
$123$ & $1/(xy)+1/(ay)+1/(bx)+1/(aS)+1/(bS)$\\[2pt]
$132$ & $-(S+x)/(bxS)$\\[2pt]
$213$ & $-(S+y)/(ayS)$\\[2pt]
$231$ & $-1/(bS)$\\[2pt]
$312$ & $-1/(aS)$\\[2pt]
$321$ & $(a+b)/(abS)$\\
\bottomrule
\end{tabular}

\end{table}

The inverse propagator after a first external leg $i$ is denoted by $x_i=(\ell+k_i)^2$, and the inverse propagator after two external legs is denoted by $y_{ij}=(\ell+k_i+k_j)^2$. The fourfold face gives
\begin{align}
 x_1&=x,&x_2&=y-x-a,&x_3&=-y-b-c,
 \label{eq:single-rhos}\\
 y_{12}&=y,&y_{13}&=x-y-b,&y_{23}&=-x-a-c.
 \label{eq:double-rhos}
\end{align}
The three first-edge pairings are
\begin{align}
 m_{123}x_1-m_{213}x_2&=f,\notag\\
 m_{132}x_1-m_{312}x_3&=-f,\notag\\
 m_{231}x_2-m_{321}x_3&=f.
 \label{eq:first-pairings}
\end{align}
The three second-edge pairings are
\begin{align}
 m_{123}y_{12}-m_{132}y_{13}&=f,\notag\\
 m_{213}y_{12}-m_{231}y_{23}&=-f,\notag\\
 m_{312}y_{13}-m_{321}y_{23}&=f.
 \label{eq:second-pairings}
\end{align}
Substituting Table~\ref{tab:weights} verifies each equality in the rational function field. The exchange oddness of $Q$ and $CQ$ then produces precisely $\cyc_{123}(Q+CQ)$, giving Eq.~\eqref{eq:fourfold-map}.

\subsection{Monomial indexing and the full functional}
The functional was found using modular linear systems; the verification below establishes it directly over characteristic zero. The displayed weights have primitive integer normalization.

Let $p(i,j)$ be the zero-based position of $(i,j)$ in the lexicographic ordering of the fifteen pairs $1\le i<j\le6$. Let $a_1<a_2<a_3<a_4$ be the unpaired legs. The monomial
\begin{equation}
 m_{ij;\bm c}=(\ee_i\cdot\ee_j)
                     \prod_{r=1}^{4}z_{a_r,c_r}
 \label{eq:indexed-monomial}
\end{equation}
has index
\begin{equation}
 \nu=625\,p(i,j)+125c_1+25c_2+5c_3+c_4.
 \label{eq:row-index}
\end{equation}
Thus $\nu$ ranges from 0 to 9374. Table~\ref{tab:certificate} gives the complete support of $w$, the polarization pairs and mixed choices, and the corresponding residual coefficients. All unlisted weights vanish. Together with Eq.~\eqref{eq:row-index}, it specifies the functional and its pairing with the target.

Each residual coefficient has the form
\begin{equation}
 [\mathcal R_*(\Ds)]_\nu=v_\nu+(\Ds-12)t_\nu.
 \label{eq:table-expansion}
\end{equation}
The exact sums over the displayed support are
\begin{equation}
 \sum_\nu w_\nu v_\nu=-480,\qquad
 \sum_\nu w_\nu t_\nu=-48.
 \label{eq:table-sums}
\end{equation}
These are characteristic-zero sums of rational numbers.

The verification of the homogeneous identity uses the expansion of $4\mathcal U\Pi$ into the 24 signed affine actions of Appendix~\ref{app:coordinates}. Each action relabels a raw monomial's polarization pair and transforms its four mixed contractions, which are then expressed in the coordinates of Table~\ref{tab:coordinates}. Pairing with the weights of Table~\ref{tab:certificate} gives zero on all 9375 monomials. Equivalently, pulling back $w$ under the 24 signed actions and summing the resulting coefficient arrays gives zero on every raw monomial. We evaluated this sum by exact integer arithmetic, with multiplication by $6^4$ clearing all routing denominators.

\begin{table}[p]
\caption{Complete support of the integer functional $w$ and the coefficients in Eq.~\eqref{eq:table-expansion}. The pair $ij$ denotes $(\epsilon_i\cdot\epsilon_j)$, and the four digits $\bm c$ specify the mixed choices of the unpaired legs in increasing order. The row index is zero based. The entries are ordered from top to bottom in the left block and then in the right block; all unlisted weights vanish. Both residual columns are normalized by $f$ as in Eq.~\eqref{eq:residual-def}.}
\label{tab:certificate}
\footnotesize
\setlength{\tabcolsep}{3pt}
\renewcommand{\arraystretch}{0.92}
\centering
\begin{tabular}{rccrrr@{\hspace{2em}}rccrrr}
\toprule
$\nu$ & $ij$ & $\bm c$ & $w_\nu$ & $v_\nu$ & $t_\nu$ & $\nu$ & $ij$ & $\bm c$ & $w_\nu$ & $v_\nu$ & $t_\nu$\\
\midrule
7545 & 45 & 0140 & 6 & $-25/6$ & $0$ & 136 & 12 & 1021 & -108 & $-118/9$ & $-32/9$\\
7540 & 45 & 0130 & -6 & $5/3$ & $0$ & 196 & 12 & 1241 & 108 & $-272/27$ & $-64/27$\\
7509 & 45 & 0014 & -6 & $13/6$ & $0$ & 259 & 12 & 2014 & -36 & $-353/18$ & $4/9$\\
7516 & 45 & 0031 & 6 & $0$ & $0$ & 190 & 12 & 1230 & -36 & $70/3$ & $8/3$\\
1301 & 14 & 0201 & 12 & $10/3$ & $0$ & 175 & 12 & 1200 & 6 & $140/3$ & $16/3$\\
1319 & 14 & 0234 & 36 & $29/12$ & $0$ & 258 & 12 & 2013 & 36 & $-110/9$ & $8/9$\\
1305 & 14 & 0210 & -12 & $-5/6$ & $0$ & 211 & 12 & 1321 & 216 & $-71/9$ & $-16/9$\\
7521 & 45 & 0041 & -6 & $0$ & $0$ & 169 & 12 & 1134 & 216 & $-349/36$ & $-2/9$\\
1316 & 14 & 0231 & -36 & $5/3$ & $0$ & 200 & 12 & 1300 & -6 & $73$ & $8$\\
1320 & 14 & 0240 & 12 & $-16/3$ & $0$ & 202 & 12 & 1302 & 18 & $-83/2$ & $-4$\\
1304 & 14 & 0204 & -12 & $29/6$ & $0$ & 172 & 12 & 1142 & 108 & $-7/9$ & $-8/9$\\
1308 & 14 & 0213 & 36 & $5/18$ & $0$ & 138 & 12 & 1023 & 72 & $40/9$ & $-16/9$\\
1425 & 14 & 1200 & 6 & $1/6$ & $0$ & 179 & 12 & 1204 & -54 & $-286/9$ & $-8/9$\\
1428 & 14 & 1203 & -36 & $-1/18$ & $0$ & 132 & 12 & 1012 & -72 & $47/6$ & $-4/3$\\
1725 & 14 & 3400 & -6 & $-8/3$ & $0$ & 173 & 12 & 1143 & -216 & $58/27$ & $-16/27$\\
1323 & 14 & 0243 & -36 & $-20/9$ & $0$ & 160 & 12 & 1120 & -36 & $61/9$ & $8/9$\\
1335 & 14 & 0320 & -6 & $-16/3$ & $0$ & 153 & 12 & 1103 & 18 & $-124/9$ & $-8/9$\\
1327 & 14 & 0302 & 6 & $8$ & $0$ & 134 & 12 & 1014 & 72 & $353/18$ & $-4/9$\\
1390 & 14 & 1030 & -6 & $0$ & $0$ & 197 & 12 & 1242 & 108 & $-38/9$ & $-16/9$\\
1375 & 14 & 1000 & 1 & $0$ & $0$ & 149 & 12 & 1044 & 36 & $130/9$ & $-16/9$\\
1475 & 14 & 1400 & 6 & $0$ & $0$ & 271 & 12 & 2041 & 36 & $344/9$ & $64/9$\\
1514 & 14 & 2024 & 36 & $-29/18$ & $0$ & 142 & 12 & 1032 & -72 & $-33/2$ & $-4$\\
1522 & 14 & 2042 & -36 & $4/3$ & $0$ & 130 & 12 & 1010 & -6 & $79/3$ & $8/3$\\
1527 & 14 & 2102 & -36 & $1/12$ & $0$ & 178 & 12 & 1203 & 18 & $-194/9$ & $-16/9$\\
1507 & 14 & 2012 & 36 & $-5/12$ & $0$ & 207 & 12 & 1312 & 216 & $37/12$ & $-2/3$\\
1354 & 14 & 0404 & 6 & $8$ & $0$ & 209 & 12 & 1314 & -216 & $215/36$ & $-2/9$\\
1511 & 14 & 2021 & -36 & $-10/9$ & $0$ & 214 & 12 & 1324 & -216 & $-1/18$ & $-4/9$\\
8132 & 46 & 0012 & 6 & $-2/3$ & $0$ & 501 & 12 & 4001 & 6 & $0$ & $0$\\
8170 & 46 & 0140 & 6 & $11/3$ & $0$ & 148 & 12 & 1043 & -36 & $-28/9$ & $-32/9$\\
1925 & 15 & 0200 & -1 & $-4$ & $0$ & 144 & 12 & 1034 & 36 & $65/6$ & $-4/3$\\
1926 & 15 & 0201 & -6 & $-4/3$ & $0$ & 154 & 12 & 1104 & -54 & $-493/18$ & $-4/9$\\
1902 & 15 & 0102 & 6 & $-8$ & $0$ & 171 & 12 & 1141 & 108 & $-100/27$ & $-32/27$\\
1901 & 15 & 0101 & 6 & $-8$ & $0$ & 183 & 12 & 1213 & 216 & $83/27$ & $-8/27$\\
1920 & 15 & 0140 & -6 & $0$ & $0$ & 176 & 12 & 1201 & 36 & $-334/9$ & $-32/9$\\
1905 & 15 & 0110 & 6 & $0$ & $0$ & 1351 & 14 & 0401 & -6 & $0$ & $0$\\
1904 & 15 & 0104 & -6 & $-8$ & $0$ & 198 & 12 & 1243 & -216 & $44/27$ & $-32/27$\\
1900 & 15 & 0100 & 2 & $0$ & $0$ & 266 & 12 & 2031 & -36 & $89/3$ & $16/3$\\
1877 & 15 & 0002 & 2 & $0$ & $0$ & 201 & 12 & 1301 & -18 & $-161/3$ & $-16/3$\\
1886 & 15 & 0021 & 6 & $0$ & $0$ & 278 & 12 & 2103 & -36 & $124/9$ & $8/9$\\
2100 & 15 & 1400 & -6 & $16/3$ & $0$ & 210 & 12 & 1320 & 36 & $73/3$ & $8/3$\\
1977 & 15 & 0402 & -6 & $0$ & $0$ & 161 & 12 & 1121 & -216 & $-95/27$ & $-16/27$\\
1975 & 15 & 0400 & -1 & $-16$ & $0$ & 135 & 12 & 1020 & -30 & $158/3$ & $16/3$\\
167 & 12 & 1132 & -216 & $-21/4$ & $-2/3$ & 150 & 12 & 1100 & -6 & $61/3$ & $8/3$\\
129 & 12 & 1004 & 12 & $-79/3$ & $-8/3$ & 152 & 12 & 1102 & 36 & $-103/6$ & $-4/3$\\
163 & 12 & 1123 & 216 & $-34/27$ & $-8/27$ & 177 & 12 & 1202 & -36 & $-88/3$ & $-8/3$\\
191 & 12 & 1231 & -216 & $-107/9$ & $-16/9$ & 139 & 12 & 1024 & 36 & $137/9$ & $-8/9$\\
182 & 12 & 1212 & -216 & $16/9$ & $-4/9$ & 147 & 12 & 1042 & 36 & $-62/3$ & $-16/3$\\
194 & 12 & 1234 & 216 & $-71/9$ & $-4/9$ & 1979 & 15 & 0404 & 6 & $-16/3$ & $0$\\
\bottomrule
\end{tabular}

\end{table}

% Keep the full-page coefficient table with Appendix C.
\clearpage

\subsection{An additional rational-point check}
The second scalar point is
\begin{equation}
 \begin{aligned}
 &(s_{12},s_{13},s_{14},s_{15},s_{23},s_{24},s_{25},s_{34},s_{35})\\
 &\hspace{10pt}=(3,7,11,13,5,17,19,23,29),\\
 &(\rho_0,\ldots,\rho_5)=(0,31,37,0,0,0).
 \end{aligned}
 \label{eq:alternate-point}
\end{equation}
The same coefficient-extraction rule and physical recursion reproduce all $v_\nu$ and $t_\nu$ of Table~\ref{tab:certificate} after division by the corresponding $f$. This agreement is a check of the calculation at another regular specialization. No independence of these coefficients from all scalar kinematics is needed for the argument.

\section{An additional exact reconstruction}
\label{app:independent}
As an additional check of the obstruction, we reconstructed the cut equations using a different maximal-cut polynomial and a specialization of the external polarizations. The resulting necessary system gives the same incompatibility. We describe the construction and its principal exact results below.

\subsection{A different maximal-cut polynomial}
This construction starts from field-strength operators rather than the particular fivefold solution $N_1$. Exact inverse-problem formulations have also been applied to five-gluon numerators~\cite{Mai2026}. For the route in Eq.~\eqref{eq:qroute}, the scalar factors and field-strength operators are defined by
\begin{equation}
 a_i=\ee_i\cdot q_{i-1},\qquad
 F_i=k_i\otimes\ee_i-\ee_i\otimes k_i,\qquad
 B_i=a_iI-F_i.
 \label{eq:independent-B}
\end{equation}
On the maximal cut, these operators satisfy
\begin{equation}
 B_iq_i=a_iq_{i-1},\qquad
 q_{i-1}^{\flat}B_i=a_iq_i^{\flat}.
 \label{eq:independent-endpoints}
\end{equation}
The oriented open cubic operator is $-2B_i$ on the physical quotient. The two nonphysical trace contributions are again the product of the $a_i$. In the stripped vertex normalization of this reconstruction, the maximal-cut polynomial is
\begin{equation}
 H_6=64\left[\Tr_{\Ds}(B_1B_2B_3B_4B_5B_6)
                         -2\prod_{i=1}^{6}a_i\right].
 \label{eq:independent-H}
\end{equation}
It is polynomial, has total degree six and the required polarization multidegree, and has routed dihedral covariance. Its lower cuts are not assumed to match. The construction of this maximal-cut expression is established one-loop numerator technology~\cite{Edison2023,Cao2026}.

If a full numerator exists, its difference from $H_6$ belongs to the same maximal-cut ideal. The degree argument of Section~\ref{sec:contact} therefore again gives one inverse-propagator factor times four mixed contractions in the one-pair sector.

\subsection{Polarization restriction and the enlarged space}
For a chosen pair of labels $a<b$, the polarization specialization is
\begin{equation}
 \ee_a\cdot\ee_b=1,\qquad
 \ee_a\cdot\ell=\ee_b\cdot\ell=0,
 \label{eq:special-pair}
\end{equation}
with $\ee_a\cdot k_j=\ee_b\cdot k_j=0$ for every $j$. All other cross-polarization products vanish. For an unpaired leg $i$, the mixed contractions are
\begin{equation}
 \ee_i\cdot\ell=u_i,\qquad
 \ee_i\cdot k_a=v_i,\qquad
 \ee_i\cdot k_b=-v_i,
 \label{eq:special-mixed}
\end{equation}
and $\ee_i\cdot k_j=0$ when $j\notin\{a,b\}$. These assignments satisfy transversality and momentum conservation. The unpaired $u_i,v_i$ are independent. Polarization squares may be assigned arbitrarily, since they do not enter a multilinear numerator.

This restriction removes the zero-pair grade because the paired polarizations have no momentum contractions. It removes every higher-pair grade because only one cross-polarization product is nonzero. It therefore isolates the complete one-pair sector under every routed word. A polynomial identity for arbitrary transverse polarizations must hold on these specialized data whenever its denominators are regular.

Across an unpaired external leg, the adjacent loop gaps are identified by their contractions with every unpaired polarization. Removing the four unpaired legs leaves two components of gaps, separated by the paired legs. For each unpaired $i$, let $U_{i,0}$ and $U_{i,1}$ be the two values of $\ee_i\cdot q$ on these components. They are invertible linear reparameterizations of $u_i,v_i$ for a fixed route.

Every specialized contact correction lies in the span of
\begin{equation}
 (\ee_a\cdot\ee_b)\,\rho_j
       \prod_{i\notin\{a,b\}}U_{i,\eta_i},\qquad
 j=0,\ldots,5,\quad \eta_i\in\{0,1\}.
 \label{eq:1440}
\end{equation}
Over all pairs, this gives $15\cdot6\cdot2^4=1440$ labeled monomials. Allowing arbitrary coefficients for all of them may admit specialized polynomials that do not lift to full numerator polynomials. It thus enlarges the possible image and preserves the necessary implication from any full solution.

The dihedral action permutes the pairs, gaps, and the two gap components. Reflection negates each $U$, leaving the product of four invariant in sign. The action partitions the 1440 monomials into 126 orbits. Their orbit sums span the entire invariant enlarged space. The vector $x\in\mathbb F^{126}$ contains the orbit-sum coefficients.

\subsection{Cut equations and exact elimination}
The graph-side cut includes all ordered forests of rooted binary trees whose leaves partition each corner containing $b$ consecutive external gluons. Every nontrivial subtree contributes its external quadratic propagator; successive roots contribute the intervening uncut loop propagator. For $b=1,2,3,4$, the enumeration gives $1,2,5,14$ planar cubic trees, respectively. Expanding each root through $[A,B]=AB-BA$ determines all signed master words. Each is evaluated at the same incoming loop momentum at the corner.

The graph side evaluates the orbit sums and the fixed polynomial $H_6$. The physical side uses the full cubic and quartic recursion of Appendix~\ref{app:operators}. After subtracting the graph-side value of $H_6$, each exact evaluation gives a row linear in $x$. A positive integer clears all rational denominators in that row, including the two target columns. Writing its physical target as a constant term plus its state-count coefficient gives
\begin{equation}
 A_{\rm B}x=b_0+\Ds b_1.
 \label{eq:independent-system}
\end{equation}

We evaluated 103 equations at regular rational configurations: 80 with corner-size pattern $(2,1,1,1,1)$, 13 with pattern $(3,1,1,1)$, and 10 with pattern $(4,1,1)$. Every cut contains at least three distinct loop edges. Tadpoles and bubbles cannot change these residues, although they remain allowed in the full representation.

The matrix $A_{\rm B}$ has 103 rows and 126 numerator columns. Exact rational elimination gives rank 102 and a left-null vector $\lambda$. Direct multiplication verifies its action on all numerator columns and both target vectors. With the normalization $\lambda^Tb_1=1$, we obtain
\begin{equation}
 \begin{gathered}
 \rank A_{\rm B}=102,\qquad \lambda^TA_{\rm B}=0,\\
 (\lambda^Tb_0,\lambda^Tb_1)=(-2,1).
 \end{gathered}
 \label{eq:independent-rank}
\end{equation}
The elimination used fraction-free integer operations followed by rational back substitution. The 103 nonzero components of $\lambda$ act on the denominator-cleared equations.

We certified the rank by a nonsingular $102\times102$ minor together with the nonzero left-null vector of the full matrix. The former gives rank at least 102 and the latter gives rank at most 102, without a numerical rank decision. Contracting the target with $\lambda$ gives $0=\Ds-2$ over every characteristic-zero coefficient field, in agreement with the obstruction derived in Section~\ref{sec:obstruction}.

\subsection{Rational kinematics and consistency checks}
This reconstruction uses the external labels $1,\ldots,6$ and the incoming loop momentum $\ell=q_0$ of Section~\ref{sec:conventions}. The scalar coordinates are
\begin{equation}
 s_i=(k_i+k_{i+1})^2,\qquad
 t_i=(k_i+k_{i+1}+k_{i+2})^2,
 \label{eq:independent-st}
\end{equation}
for cyclic indices $i=1,\ldots,6$, with $t_{i+3}=t_i$. Adjacent and next-to-adjacent products are
\begin{align}
 k_i\cdot k_{i+1}&=s_i/2,\notag\\
 k_i\cdot k_{i+2}&=(t_i-s_i-s_{i+1})/2.
 \label{eq:independent-products}
\end{align}
Opposite products follow from $k_i\cdot\sum_jk_j=0$. With $\rho_i=q_i^2$ as in Eq.~\eqref{eq:rhos}, the loop products are
\begin{equation}
 \ell^2=\rho_0,\qquad
 \ell\cdot k_i=\frac{\rho_i-\rho_{i-1}}{2}
                            -\sum_{j=1}^{i-1}k_i\cdot k_j,
 \label{eq:independent-loop}
\end{equation}
where $\rho_6=\rho_0$.
Equations~\eqref{eq:special-pair} to \eqref{eq:special-mixed} fix the remaining relevant Gram entries.

For the Gram determinant checks, the diagonal polarization entries are fixed as
\begin{equation}
 \ee_a^2=\ee_b^2=0,\qquad
 \ee_i^2=1\quad(i\notin\{a,b\}).
 \label{eq:gram-diagonal}
\end{equation}
The paired labels are those of Eq.~\eqref{eq:special-pair}. We checked that all 103 twelve-vector Gram matrices are nondegenerate with this prescription, so each has a complex realization in dimension twelve. A representative evaluation has cut partition $(4,1,1)$ and
\begin{equation}
 \begin{aligned}
  (s_1,\ldots,s_6)&=(28,23,19,33,14,27),\\
  (t_1,t_2,t_3)&=(88,77,84),\\
  (\rho_0,\ldots,\rho_5)&=(0,149,118,126,0,0),\\
  (a,b)&=(1,4).
 \end{aligned}
 \label{eq:independent-example}
\end{equation}
For the unpaired labels $2,3,5,6$, the corresponding $(u_i,v_i)$ are $(1,39)$, $(27,49)$, $(22,8)$, and $(39,47)$. The Gram determinant is
\begin{equation}
 \det G=-\frac{3079605846289223}{64}\ne0.
 \label{eq:independent-det}
\end{equation}
The diagonal prescription fixes this determinant but does not affect the multilinear cut coefficients, which contain no polarization self-contractions.

We reconstructed every physical cut equation from the tree recursion and performed the checks summarized in Table~\ref{tab:checks}. As a lower-point control, we evaluated the local four-gluon numerator of Cao et al.~\cite{Cao2026} on maximal, triple, and double cuts at three rational kinematic configurations, with internal vector-state dimensions twelve and thirteen. This tests the normalization and physical cut reconstruction against an explicit local representation.

\begin{table}[htbp]
\caption{Checks performed in the additional reconstruction. All comparisons use exact rational arithmetic; no numerical tolerance is introduced.}
\label{tab:checks}
\centering
\begin{tabular}{lr}
\toprule
Check & Number\\
\midrule
Reconstructed physical cut equations & 103\\
Nonzero twelve-vector Gram determinants & 103\\
Direct-projector cut comparisons & 206\\
External Ward replacements & 24\\
Maximal-cut comparisons & 4\\
Routed dihedral comparisons & 48\\
Published four-point comparisons & 18\\
\bottomrule
\end{tabular}

\end{table}

\begingroup
\interlinepenalty=10000
\bibliographystyle{JHEP}
\bibliography{references}
\endgroup
\end{document}